\documentclass[doublecol,figures]{epl2}

\usepackage{ifpdf}
\newcommand{\ra}{\rightarrow}
\newcommand{\be}{\begin{equation}}
\newcommand{\ee}{\end{equation}}
\newcommand{\ba}{\begin{array}}
\newcommand{\ea}{\end{array}}
\newcommand{\pa}{\partial}
\newcommand{\bi}{\bibitem}

\revision

\title{Glueballs and the Pomeron}

\author{M. N. Sergeenko}
\institute{ The National Academy of Sciences
of Belarus - Pr. Nezavisimosti, 70, Minsk, 220072, Belarus\\
 State University of Transport - Kirov Street, 34, Gomel,
246653, Belarus\footnote {The present address. Email:
msergeen@usa.com}}

\pacs{12.39.Mk}{Glueball and nonstandard multi-quark/gluon states}
\pacs{12.39.Pn}{Potential models}
\pacs{12.40.Nn}{Regge theory, duality}
\pacs{12.40.Yx}{Hadron mass models and calculations}

\abstract{ Glueballs are considered to be bound states of
constituent gluons. Relativistic wave equation for two massive
gluons interacting by the funnel-type potential is analyzed.
Using two exact asymptotic solutions of the equation, we derive
an interpolating mass formula and calculate glueball masses in
agreement with the lattice data. We obtain the complex non-linear
Pomeron trajectory, $\alpha_P(t)$, in the whole region of $t$.
The real part of the trajectory corresponds to the soft Pomeron,
parameters of which are found from the fit of recent HERA data.}

\begin{document}
\maketitle

\section{Introduction}

Quantum Chromodynamics allows for an effective description of hadrons as
bound states of constituent particles. The classification of baryons and
mesons with the quark hypothesis is a first historical example. There
should exist a way to connect the constituent approaches to QCD. This
longstanding problem is far from being completely solved.

A good test of our understanding of the nonperturbative (NP)
aspects of QCD is to study particles where the gauge field plays a
more important dynamical role than in the standard hadrons. QCD
allows the existence of purely gluonic bound states, glueballs
\cite{KleZa,BBMS,MaKoVe}, but no firm experimental discovery of
such states has been obtained yet. Glueballs are particles whose
valence degrees of freedom are gluons. An important theoretical
achievement in this field has been the computation of the glueball
spectrum in lattice QCD \cite{Latt}.

The glueball spectrum has also been computed by using effective
approach like potential model \cite{MaSeSi,KaiSi,BraSe}.
The potential model is very successful to describe bound states
of quarks. It is also a possible approach to study glueballs
\cite{BraSe,MaBuSe}. Recent results in the physics of glueballs
have been reviewed in \cite{MaKoVe}.

At the present, one of open topics in hadron physics is the
relation between glueballs and the Pomeron. The Pomeron is the
Pomeranchuk trajectory ($P$ trajectory) \cite{DoDoLa}. In the many
high energy reactions with small momentum transfer the soft Pomeron
exchange, gives the dominant contribution \cite{DLPW}.

Recent small $-t$ ZEUS and H1 data for exclusive $\rho$ and $\phi$
photoproduction \cite{ZEUS,HERA} point out that the $P$ trajectory
is rather \emph{non-linear}. The data have been explained by adding in a
hard Pomeron contribution, whose magnitude is
calculated from the data for exclusive $J/\Psi$ photoproduction
\cite{DL,DLZ}. The ZEUS, H1 as well as CDF data on $p\bar p$
elastic scattering data have also been analyzed by using the non-linear
$P$ trajectory \cite{FiJPP}. But non-linearity of the \emph{P} 
trajectory is still an open question. The amount of non-linearity 
is unknown.

There has been a long-standing speculation that the physical
particles on the $P$ trajectory might be glueballs
\cite{MaKoVe,KleZa}. In this work we take a picture where the
$t$-channel Pomeron is dual to glueballs, i.e., purely gluonic
bound states in the $s$ channel. We use the potential approach,
which is the natural framework for studying the Regge trajectories
and their properties \cite{Collin,CollKear}. The potential model,
which is so successful to describe bound states of quarks, is also
a possible approach to study glueballs \cite{BraSe,HoLW}. We derive an
interpolating mass formula for glueball masses and analytic expression
for the $P$ trajectory, $\alpha_P(t)$, in the whole region of
variable $t$. Calculation results are compared with the lattice data.

\section{Glueballs}

Glueballs are purely gluonic bound states allowed by QCD.
At the present, there is the understanding of the deep relation
between the properties of the glueball states and the structure
of the QCD vacuum. The basic idea is that the vacuum is filled
with $J^{PC}=0^{++}$ transverse electric glueballs which form a
negative energy condensate \cite{DonJo}.

Two gluons in a color singlet state have always positive charge
conjugation ($C=+$). The lightest glueballs, which have $C=+$,
can be successfully modeled by a two-gluon system in which the
constituent gluons are massless helicity-1 particles
\cite{ViMatBS}. The proper inclusion of the helicity degrees of
freedom dramatically improves the compatibility between lattice
QCD and potential models \cite{ViMatBS}.

The modern development in glueball spectroscopy from various
perspectives has been discussed in \cite{MaKoVe}. At the present,
several candidates for the low mass glueballs with quantum numbers
$J^{PC}=0^{++}$, $2^{++}$, $0^{-+}$ and $1^{--}$ are under
discussion.

Glueball masses have been calculated by many authors.
A new method called the Vacuum Correlator Model (VCM) has been
used in \cite{Si}. In this model all nonperturbative and perturbative
dynamics of quarks and gluons is universally described by lowest
cumulants, i.e., gauge invariant correlators of the type
$\langle F_{\mu\nu}(x_1)\ldots F_{\lambda\sigma} (x_{\nu})\rangle$.
More discussions on the subject can be found in \cite{MaKoVe}.

Authors of \cite{BraSe} compared different models for glueballs.
They concluded that a semi-relativistic Hamiltonian is an essential
ingredient to handle glue states. All models analyzed used an
$SL$-basis to include spin of gluons. These arguments support the
use of an \emph{effective gluon mass} to describe the glueball
dynamics of QCD.

\section{Reggeons and the Pomeron}

There exists a conviction, that the Regge trajectories are
linear in the whole region, that is, not only in
the bound state region ($t=E^2>0$) but in the scattering region
($t<0$), too. However, one of the most crucial distinctions
between small $-t$ behavior and large $-t$ behavior of
trajectories $\alpha(t)$ involves the asymptotic form of Regge
trajectories at $-t\ra\infty$.

The asymptotic behavior of the Regge trajectories at $-t\ra\infty$
has been discussed by many authors \cite{Bro,Ans,Pet}. The
constituent interchange model (CIM) \cite{Bro} results in the
prediction for the large $-t$ behavior of $\rho$ trajectory
\be
\alpha_{\rho}(t)=-1, ~~~t\ra-\infty, \label{at1} \ee that means
the $\rho$ trajectory is non-linear. General properties of the
trajectories have been considered in classical papers \cite{GriPo,DeP}.

There is a renewed interest in the studies of the dynamics of the
Regge trajectories \cite{FiPaSr}. The conception of linear Regge
trajectories is not consistent with experimental data and
expectations of perturbative QCD (pQCD) at large
$-t\gg\Lambda_{QCD}$ \cite{Bro}. In the experiment far more
complicated behavior of the $\rho$ meson trajectory,
$\alpha_\rho(t)$, was discovered; the $\rho$ trajectory flattens
off at about $-0.6$ or below.

Regge trajectories with the same asymptotic behavior for all
leading $S=1$ meson and quarkonium Regge trajectories were
obtained in our Refs. \cite{SeA,SeZ}. The calculated effective
$\rho$ Regge trajectory matching the experimental data on the
spectrum of the $\rho$ trajectory as well as those on the
charge-exchange reaction $\pi^-p\ra\pi^0n$ at $t<0$. The trajectory
deviates considerably from a linear in the space-like region and
is asymptotically linear in the time-like region, matching nicely
between the two.

The Pomeron is the highest-lying Regge trajectory.
In the many high energy reactions with small momentum transfer
the Pomeron exchange, gives the dominant contribution \cite{DLPW,LySe}.
The classic soft Pomeron is constructed from multi-peripheral
hadronic exchanges. It is usually believed that the soft $P$
trajectory is a linear function,
\be
\alpha_P(t)=\alpha_P(0)+\alpha_P^\prime (0)t,
\label{alPt} \ee
where the intercept $\alpha_P(0)=1$ and the slope
$\alpha_P^\prime(0)=0.25$ GeV$^{-2}$. These fundamental parameters
are the most important in high-energy hadron physics. Usually,
they are determined from experiment in hadron-hadron collisions.

To explain the rising hadronic cross sections at high energies,
the classic soft Pomeron was replaced by a soft supercritical
Pomeron with an intercept $\alpha_P(0)\simeq 1.08$. What is
the Pomeron by definition?

The Pomeron is the vacuum exchange contribution to scattering at
high energies at leading order in $1/N_c$ expansion. In gauge
theories with string-theoretical dual descriptions, the Pomeron
emerges unambiguously. In the QCD framework the Pomeron can be
understood as the exchange of at least two gluons in a color
singlet state \cite{LowN}. The pQCD approach to the Pomeron,
the Balitski\v{i}-Fadin-Kuraev-Lipatov (BFKL) Pomeron, has been
discussed in \cite{Fa,Lip}. The Pomeron can also be associated
with a reggeized massive graviton \cite{SofHardP}.

The approximate linearity (\ref{alPt}) is true only in
a small $-t$ region. The ZEUS, H1 as well as CDF data on
$p\bar p$ elastic scattering data have also been analyzed
by using the non-linear $P$ trajectory \cite{FiJPP}.
Important theoretical results have been obtained in
\cite{ErSch1,ErSch2,ErSch3}. The results imply that the effective
\emph{P} trajectory flattens for $-t > 1$ GeV$^2$ that is evidence
for the onset of the perturbative 2-gluon Pomeron. These results
may shed some light on the self-consistency of recent measurements
of hard-diffractive jet production cross sections in the UA8,
CDF and HERA experiments.

A further analysis \cite{ErSch2} of inelastic diffraction data at
the ISR and SPS-Collider confirms the relatively flat $s$-independent
\emph{P} trajectory in the high $-t$ domain, $1 < -t < 2$ GeV$2$,
reported by the UA8 Collaboration. This suggests a universal
fixed \emph{P} trajectory at high $-t$. It was shown that a triple-Regge
Pomeron-exchange parametrization fit to the data requires an
$s$-dependent (effective) \emph{P} trajectory intercept, $\alpha_P(0)$,
which decreases with increasing $s$, as expected from unitarization
(multi-Pomeron-exchange) calculations, $\alpha_P(0)=1.10$ at
the lowest ISR energy, $1.03$ at the SPS-Collider and perhaps smaller
at the Tevatron. In \cite{ErSch3} a new $\gamma^*p$/$p\bar p$
factorization test in diffraction, valid below $Q^2$ about $6$ GeV$^2$
has been investigated. The apparent factorization breakdowns are likely
due to the different effective \emph{P} trajectories in $ep$ and $pp$
interactions.

The issue of soft and hard Pomerons has been discussed extensively
in the literature \cite{Fa,Lip,DuKaSi}. Both the IR (soft) Pomeron
and the UV (BFKL) Pomeron are dealt in a unified single step.
On the basis of gauge/string duality, the authors describe
simultaneously both the BFKL regime and the classic Regge regime.
The problem was reduced to finding the spectrum of a single
$j$-plane Schr\"odinger operator. The results agreed with expectations
for the BFKL Pomeron at negative $t$, and with the expected glueball
spectrum at positive $t$, but provide a framework in which they are
unified.

A model for the Pomeron has been put forward by Landshoff and
Nachtmann where it is evidenced the importance of the QCD NP
vacuum \cite{DoDoLa}. The current data is compatible with a smooth
transition from a soft to a hard Pomeron contribution which can
account for the rise of $\sigma_{tot}$ with $s$. If soft and BFKL
Pomeron have a common origin, the discontinuity across the cut in
the $\alpha_P(t)$ plane must have a strong $t$ dependence that points
out on non-linearity of the $P$ trajectory \cite{Bjor}.

There are currently no any realistic theoretical estimations of
the $P$ trajectory. The behavior of the trajectory $\alpha_P(t)$
in the whole region is unknown. Below, in this work we reproduce
the $P$ trajectory in the whole region and calculate its
parameters $\alpha_P(0)$ and $\alpha^\prime_P(0)$.

\section{The Pomeron trajectory}

Let us consider the picture in which the physical particles on
the $P$ trajectory are glueballs, i.e., purely gluonic bound
states of massive gluons \cite{MaKoVe,KleZa}. The $P$ trajectory
can be obtained by similar way as the reggeon ones \cite{SeZ,SeA}.
We consider clueballs as relativistic two-gluon bound systems.
The question arises: what is the potential of $gg$ interaction?

The potential is nonrelativistic concept. Nevertheless, the
potential is successfully used in many relativistic models.
We do not know the QCD potential in the whole region. It is
generally agreed that, in pQCD, as in QED the essential
interaction at small distances is instantaneous Coulomb exchange;
in QCD, it is $qq$, $qg$, or $gg$ Coulomb scattering \cite{Bjor}.
For large distances, from lattice-gauge-theory computations
\cite{BaSh} follows that the potential is an approximately linear,
$V_L(r)\simeq\sigma r$, at $r\ra\infty$, where $\sigma\simeq
0.15$\,GeV$^2$ is the string tension.

In the model of \cite{Si} all dependence on gluonic fields $\bar
A_\mu$ is contained in the adjoined Wilson loop $\langle
W_{adj}(C)\rangle$, where the closed contour $C$ runs over
trajectories $z_\mu(\sigma)$ and $\bar z_\mu(\sigma)$ of both
gluons. Gluons are linked by an adjoint string. The adjoint
string tension $\sigma_a=9\sigma/4$ is expressed in terms of the
well-known fundamental string tension $\sigma$ for mesons through
the Casimir scaling hypothesis. Using typical values for the
parameters, $\sigma=0.15$ GeV$^2$ and $\alpha_s=0.4$ for the strong
coupling, this model encodes the essential features of glueballs.
More discussions on the subject can be found in \cite{MaKoVe}.

In the adjoined and fundamental representations, the final
form of interaction of two gluons is \cite{Si}:
\be
V_{gg}(r) =-\frac{\alpha_a}r +\sigma_a r -C_0, \label{Vadj} \ee
where $\alpha_a \equiv\alpha^{adj}=3\alpha_s^{fund}$, $\sigma_a
\equiv\sigma^{adj}$= $\frac 94\sigma^{fund}$; $\alpha_s^{fund}$ is
the strong coupling, $\sigma^{fund}\equiv\sigma\simeq
0.15$\,GeV$^2$ is the string tension, and $C_0$ is the arbitrary
parameter.

In hadron physics, the nature of the potential is very important.
There are normalizable solutions for scalarlike potentials, but
not for vectorlike \cite{Su}. The effective interaction has to be
scalar in order to confine particles (quarks and gluons) \cite{Su}.

To reproduce the $P$ trajectory we need to obtain an analytic
expression for the squared gluonium mass, $E^2$. For this, we
solve the relativistic semi-classical wave equation, which for
two interacting gluons of equal masses $\mu_1=\mu_2=\mu_0$ is
\cite{SeM,SeRe}
\be -\left(\frac{\pa^2}{\pa r^2} +\frac 1{r^2}
\frac{\pa^2}{\pa\theta^2} +\frac 1{r^2\sin^2\theta}
\frac{\pa^2}{\pa\varphi^2}\right)\tilde\psi(\vec r)=
p^2(E,r)\tilde\psi(\vec r) \label{eq}, \ee
where $p^2(E,r)={E^2/4 - [\mu_0+V_{gg}(r)]^2}$.

The correlation of the function $\tilde{\psi}(\vec r)$ with the wave
function ${\psi}(\vec r)$ in case of the spherical coordinates is
given by the relation $\tilde{\psi}(\vec r)=\sqrt{det\,g_{ij}}
\psi(\vec r)$, which follows from the identity: $\int\mid \psi(\vec
r)\mid^2d^3\vec r$ $\equiv \int\mid\psi(\vec r)$ $\mid^2det\,g_{ij}
dr\,d\theta\,d\varphi $ $= 1$, where $g_{ij}$ is the metric tensor
(det$\,g_{ij}=r^2\sin\,\theta$ for the spherical coordinates).

Relativistic wave equations are usually solved in terms of special
functions, with the help of specially developed methods or
numerically. However, almost together with quantum mechanics, the
appropriate method to solve the wave equation has been developed; it
is general simple for all the problems, and its correct application
results in the exact energy eigenvalues for all solvable potentials.
This is the phase-integral method which is also known as the WKB
method \cite{Shi,Fro}.

It is hard to find the exact analytic solution of equation
(\ref{eq}) for the potential (\ref{Vadj}). But we can find exact
analytic solutions for two asymptotic limits of the potential
(\ref{Vadj}), i.e. for the Coulomb and linear potentials,
separately \cite{SeM}. The most general form of the WKB solution
and the quantization condition can be written in the complex
plane \cite{SeRe}.

The WKB quantization condition appropriate to (\ref{eq})
with the Coulomb potential is
\be
I=\oint_C\sqrt{\frac{E^2}4-\mu_0^2 + \frac{2\alpha_a\mu_0}r
 - \frac{\Lambda^2}{r^2}}=2\pi\left(n_r+\frac 12\right),
\label{FasI} \ee
where $\Lambda^2=(l+1/2)^2 +\alpha_a^2$ and a contour $C$
encloses the classical turning points $r_1$ and $r_2$. Using the
method of stereographic projection, we should exclude the
singularities outside the contour $C$, i.e. at $r=0$ and $\infty$.
Excluding these infinities we have, for the integral (\ref{FasI}),
\be
I = 2\pi(\alpha_a \mu_0/\sqrt{-E^2/4+\mu_0^2}-\Lambda),
\ee
and for the energy eigenvalues this gives \cite{SeRe},
\be
E_n^2 =4\mu_0^2\left[1
-\frac{\alpha_a^2}{\left(n_r+1/2 +\Lambda\right)^2}\right]
\label{ECoul}. \ee
Eigenvalues (\ref{ECoul}) are exact, for the Coulomb potential.
Analogously, we obtain, for the linear potential \cite{SeRe},
\be E_n^2
=8\sigma_a\left(2n_r+l -\alpha_a +\frac 32\right). \label{Elin}
\ee
For small distances, where the Coulomb type contribution
dominates, the effective strong coupling, $\alpha_a$, is a small
value and (\ref{ECoul}) can be rewritten in the simpler form \be
E_n^2 \simeq 4\mu_0^2\left[1
-\frac{\alpha_a^2}{(n_r+l+1)^2}\right]. \label{ECoS} \ee

To find gluonium energy eigenvalues (glueball masses) we use the
same approach as in \cite{SeA,SeZ}, i.e., we derive an
interpolating mass formula for $E^2_n$, which satisfies both of
the above constraints (\ref{Elin}) and (\ref{ECoS}). To derive
such a formula, the two-point Pad\'e approximant can be used
\cite{Bak},
\be [K/N]_f(z)
=\frac{\sum_{i=0}^Ka_iz^i}{\sum_{j=0}^Nb_jz^j}, \label{Pappr} \ee
with $K=3$ and $N=2$. We use $K=3$ and $N=2$ because this is 
a simplest choice to satisfy the two asymptotic limits above.

The result is the interpolating mass formula \cite{SeZ,SeA},
\be E_n^2=8\sigma_a\left(2n_r+l +\frac 32
-\alpha_a\right) -\frac{4\alpha_a^2\mu_0^2}{(n_r+l+1)^2}
+4\mu_0^2. \label{Einter} \ee Expression  (\ref{Einter}) is an
Ansatz [as the potential (\ref{Vadj})], which is based on two
asymptotic expressions (\ref{Elin}) and (\ref{ECoS}). Formula
(\ref{Einter}) and its derivation are rather simple; this allows
us to get an {\it analytic} expression for the $P$ trajectory,
$\alpha_P(t)$, in the whole region.

Transform (\ref{Einter}) into the cubic equation for the angular
momentum $l$,
\be l^3 + c_1(t)l^2 +c_2(t)l +c_3(t)=0, \label{JE3}\ee
where $c_1(t)=2\tilde n +\lambda(t)$, $c_2(t)={\tilde n}^2
+2\tilde n\lambda(t)$, $c_3(t)={\tilde n}^2\lambda(t)
-\alpha_a^2\mu_0^2/2\sigma_a, \tilde n=n_r+1$, $\lambda(t)=2\tilde
n -1/2 -\alpha_a +(4\mu_0^2 -t)/8\sigma_a$. Equation (\ref{JE3})
has three (complex in general case) roots: $l_1(t)$, $l_2(t)$, and
$l_3(t)$. The real part of the first root, ${\rm Re}\,l_1(t)$,
gives the $P$ trajectory, \be \alpha_P(t)=\left\{\begin{array}{lc}
f_1(t) +f_2(t) -c_1(t)/3,& Q(t)\ge 0; \\
2\sqrt{-p(t)}\cos\left[\beta(t)/3\right] -c_1(t)/3,& Q(t)<0,
\label{alt}
\end{array}
\right. \ee where \\ \\
$f_1(t)=\sqrt[3]{-q(t)+\sqrt{Q(t)}}$,\ \ \
$f_2(t)=\sqrt[3]{-q(t)-\sqrt{Q(t)}}$,
$$Q(t)=p^3(t)+q^2(t),\ \ p(t)=-c_1^2(t)/9+c_2(t)/3, $$
$$q(t)=c_1^3(t)/27-c_1(t)c_2(t)/16 +c_3(t)/12,$$
$$\beta(t)=\arccos\left[-q(t)/ \sqrt{-p^3(t)}\right].$$

Expression (\ref{alt}) supports existing experimental data and
reproduces the soft $P$ trajectory in the whole region of $t$ (see
below). The corresponding parameters $\alpha_a$, $\sigma_a$ and
$\mu_0$ are listened in table 1.
\begin{center}
{{\bf Table 1.} Glueball masses and parameters of \\
the $gg$ potential (\ref{Vadj})}
\begin{tabular}{cllll}
\hline \hline
$Method$ & $J^{PC}$ & $\ \ E_n^{Gl}$ & $ Parameters$ \\
\hline\hline
I & $ 2^{++}$ & $\ 1.740$ & $\alpha_a=2.448$ - fixed\\
~&$ 3^{--}$ & $\ 2.452$ & $\sigma_a=0.338$\,GeV$^2$ - fixed\\
~&$ 4^{++}$ & $\ 2.974$ & $\mu_0=0.495$\,GeV - fixed\\
~&$ 5^{--}$ & $\ 3.408$ & ~ \\
~&$ 6^{++}$ & $\ 3.789$ & ~ \\
\hline
II & $ 2^{++}$ & $\ 1.984$ & $\alpha_a=2.276\pm 0.041$ \\
~&$ 3^{--}$ & $\ 2.689$ & $\sigma_a=0.294\pm 0.003$\,GeV$^2$ \\
~&$ 4^{++}$ & $\ 3.164$ & $\mu_0=0.968\pm 0.147$\,GeV \\
~&$ 5^{--}$ & $\ 3.549$ & ~ \\
~&$ 6^{++}$ & $\ 3.884$ & ~ \\
\hline
III & $ 2^{++}$ & $\ 1.695$ & $\alpha_a=2.442\pm 0.044$\\
~&$ 3^{--}$ & $\ 2.393$ & $\sigma_a=0.323\pm 0.071$\,GeV$^2$ \\
~&$ 4^{++}$ & $\ 3.904$ & $\mu_0=0.478\pm 0.084$\,GeV \\
~&$ 5^{--}$ & $\ 3.330$ & ~ \\
~&$ 6^{++}$ & $\ 3.703$ & ~ \\
\hline\hline
\end{tabular}
\end{center}
We calculate glueball masses and the $P$ trajectory for three
different sets of parameters (methods):
\begin{figure}[h]
\onefigure[scale=1.5,width=0.45\textwidth]{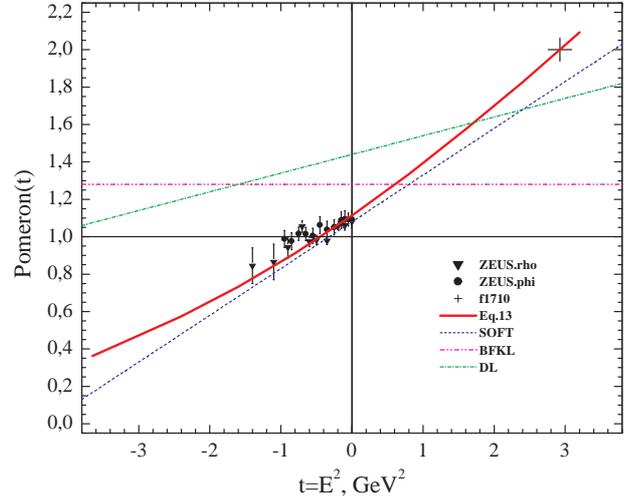}
\caption{\label{fig:Pom.eps}
The effective Pomeron trajectory. Solid curve is the trajectory
(\protect\ref{alt}) with the parameters found from the fit of
combined HERA $\rho$ (triangles) and $\phi$ (circles) data
\protect\cite{HERA}, and $2^{++}$ glueball candidate $f_0(1710)$
(cross) \protect\cite{Kirk}. Other lines show the classic soft,
BFKL, and Donnachie-Landshoff hard Pomerons.}
\end{figure}

I) the parameters are
fixed as in \cite{Si}: $\alpha_a=3\alpha_s=2.448$,
$\sigma_a=9\sigma/4=0.338$\,GeV$^2$, and gluon mass $\mu_0=1.5\,
m_q=0.495$ GeV (see also \cite{Halz}) for the typical values
$\alpha_s=0.816$, string tension $\sigma=0.15$\,GeV$^2$, quark
mass $m_q=0.330$ GeV of light mesons; II) the parameters
$\alpha_a$, $\sigma_a$, and $\mu_0$ are found from the fit of HERA
data for the $P$ trajectory \cite{HERA}; III) include into the fit
the $J^{PC}=2^{++}$ glueball candidate for $m_G=1.710$ GeV
\cite{Kirk} by supposing that the glueball trajectory is the soft
$P$ trajectory. Masses of gluonium leading states, $E_n^{Gl}$,
have been calculated with the use of the interpolating mass
formula (\ref{Einter}). The methods I and III reproduce the
trajectory with the properties of the supercritical soft Pomeron.
The effective intercept and slope estimated by these two methods
are: \be \alpha_P(0)=1.09\pm 0.02, \ \ \alpha_P^\prime(0)=0.26\pm
0.03 GeV^{-2}. \label{Ppar} \ee The corresponding effective mass
of the $2^{++}$ glueball candidate is $m_G\simeq 1.74$ GeV.

Several Pomerons are shown in Fig.\,\ref{fig:Pom.eps}. Solid
line is the effective \emph{P} trajectory (\ref{alt}). The
trajectory is asymptotically linear at $t\ra\infty$ with the slope
$\alpha_P^\prime =1/(8\sigma_a)\simeq 0.38$\,GeV$^{-2}$. In the
scattering region, the trajectory flattens off at $-1$ for $t\ra
-\infty$. We see that the experimental data and simple calculations
in the framework of the potential approach support the conception
of the soft supercritical Pomeron as observed at the presently
available energies.

The perturbative (Coulomb) part of the potential (\ref{Vadj})
gives the asymptotic expression (\ref{ECoS}), inverting which we
have ($t=E^2$)
\be l(t)=-1 +\frac{\alpha_a}{\sqrt{1 -
t/(4\mu_0^2)}}, ~~~ t\rightarrow-\infty. \label{PSas} \ee
If we take into account spins of gluons with the total spin of two
interacting gluons $S=2$ (leading states), then the formula
\be
\alpha(t)=l(t) +S\label{alS} \ee gives the asymptote for the BFKL
Pomeron predicted by pQCD \cite{Lip}.

The Pomeron with such properties has been also
used to describe the HERA data on the charm structure function
$F_2^c$ \cite{HERA,ZEUSC}. It was shown that the two-Pomeron
picture (soft plus hard Pomeron) gives a very good fit to the
total cross section for elastic $J/\Psi$ photoproduction and the
charm structure function $F_2^c$ over the whole range of $Q^2=-t$
\cite{CDL}. However, the results of these experiments and the
found higher order corrections make it quite unclear what the
hard Pomeron is.

\section{Conclusion}

Glueballs are a good test of our understanding of the
nonperturbative aspects of QCD. Their existence is allowed by
QCD and the glueball spectrum has been computed in lattice QCD.

We have considered glueballs as bound states of constituent
massive gluons and investigated their properties in the framework
of the potential approach. For two-gluon system, we have analyzed
exact asymptotic solutions of relativistic wave equation with the
funnel-type potential. Using the asymptotes, we have derived the
interpolating mass formula (\ref{Einter}) and calculated the
glueball masses, which are in agreement with the lattice data.

We have considered glueballs as the physical particles on the
\emph{P} trajectory. To reproduce the $P$ trajectory, we have
derived the interpolating mass formula (\ref{Einter}) for the squared
energy eigenvalues, $E_n^2=E^2(l,n_r)$, of the \emph{gg} system.
We have calculated gluonium masses and obtained the analytic expression
(\ref{alt}) for the $P$ trajectory, $\alpha_P(t)$ in the whole region.
In the scattering region, at $-t\gg\Lambda_{QCD}$, the trajectory
flattens off at $-1$ and has asymptote $\alpha_P(t\ra
-\infty)=-1$. In bound state region, the $P$ trajectory is
approximately linear in accordance with the string model.

However, the non-linearity of trajectories is still an open question. 
The curvature of the trajectory may come from several linear trajectories, 
as Donnachie and Landshoff showed. Here we have considered one of 
possible scenario. We do not have enough experimental data to make 
final conclusion.

It is well known, that the fixed-number of particles within the
potential approach can not be used for strict relativistic
description. Strict description of the Pomeron presupposes
multiparticle system. For perturbative regime with the Pomeron
scattering, the dominant contribution comes from the BFKL Pomeron.
However, experimental data and our rather simple calculations
support the conception of the soft supercritical Pomeron as
observed at the presently available energies. The hard BFKL
Pomeron has intercept $\alpha_P(0)=1.44$. Next-to-leading
order estimates give for the BFKL intercept 1.26 to 1.3, which
is closer to the soft supercritical Pomeron.

In this paper we have not considered helicities and spin
properties of gluons. This topic has been discussed in details
somewhere else \cite{ViMatBS}. The existing data and simple
analysis performed in this work confirm the existence of the
Pomeron whose trajectory is non-linear and corresponds to the
supercritical soft Pomeron at small spacelike $t$.

In many Regge models \cite{Kaid,SeNP}, one-Pomeron exchange gives
only dominant contribution into the cross section. With energy
growth, multiple Pomeron exchanges (MPE) and sea quark
contributions become important. The MPE contributions are
important just at small $x$ and can give explanation of the small
$x$ charm production data at HERA.

Combined with the eikonal model the MPE contributions give the
correct energy dependence of total and total inelastic cross
sections \cite{Kaid}, allow to describe hard distributions of
secondary hadrons \cite{SeNP}. From this point of view, the
required hard Pomeron discussed in \cite{DL} effectively
accounts for the MPE contributions.

\acknowledgments The author thanks V.A. Petrov and V. Mathieu for
reading the paper, giving useful comments and remarks.

\end{document}